\documentclass[american,english,aps,preprint]{revtex4}
\usepackage[T1]{fontenc}
\usepackage[latin9]{inputenc}
\usepackage{color}
\usepackage{babel}
\usepackage{verbatim}
\usepackage{float}
\usepackage{amstext}
\usepackage{graphicx}
\usepackage{subscript}
\usepackage[unicode=true,pdfusetitle,
 bookmarks=true,bookmarksnumbered=false,bookmarksopen=false,
 breaklinks=false,pdfborder={0 0 1},backref=false,colorlinks=false]
 {hyperref}

\makeatletter
\@ifundefined{textcolor}{}
{%
 \definecolor{BLACK}{gray}{0}
 \definecolor{WHITE}{gray}{1}
 \definecolor{RED}{rgb}{1,0,0}
 \definecolor{GREEN}{rgb}{0,1,0}
 \definecolor{BLUE}{rgb}{0,0,1}
 \definecolor{CYAN}{cmyk}{1,0,0,0}
 \definecolor{MAGENTA}{cmyk}{0,1,0,0}
 \definecolor{YELLOW}{cmyk}{0,0,1,0}
}

\adddialect\l@ENGLISH\l@english

\makeatother

\begin{document}
\title{{\normalsize{}Characteristics of long-lived persistent spectral holes
in Eu}\textsuperscript{{\footnotesize{}3+}}{\normalsize{}:Y}\textsubscript{{\footnotesize{}2}}{\normalsize{}SiO}\textsubscript{{\footnotesize{}5}}{\normalsize{}
at 1.2~K}}
\author{{\normalsize{}René Oswald, Michael Hansen, Eugen Wiens, Alexander
Yu. Nevsky, and Stephan Schiller}}
\email{step.schiller@hhu.de}

\homepage{http://www.exphy.uni-duesseldorf.de/}

\address{Institut für Experimentalphysik, Heinrich-Heine-Universität Düsseldorf,
40225 Düsseldorf, Germany}
\begin{abstract}
Properties of persistent spectral holes (SHs) relevant for frequency
metrology have been investigated in the system Eu\textsuperscript{{\small{}3+}}:Y\textsubscript{{\small{}2}}SiO\textsubscript{{\small{}5}}(0.5\%)
at crystallographic site 1 and a temperature of $1.2\,$Kelvin. Hole
linewidths as small as 0.6~kHz have been reliably achieved. The theoretically
predicted $T^{4}$ dependence of the frequency shift with temperature
has been confirmed with high precision. The thermal hysteresis of
the SH frequency between 1.15~K and 4.1~K was measured to be less
than $6\times10^{-3}$ fractionally. After initially burning a large
ensemble of SHs, their properties were studied on long time scales
by probing different subsets at different times. SHs could still
be observed 49~days after burning if not interrogated in the meantime.
During this time, the SH linewidth increased from 4 to 5.5~kHz, and
the absorption contrast decreased from 35\% to 15\%. During a 14-day
interval the absolute optical frequencies of previously unperturbed
spectral holes were measured with respect to a GPS-monitored active
H-maser, using a femtosecond frequency comb. The fractional frequency
drift rate exhibited an upper limit of $2.3\times10^{-19}\,{\rm s}^{-1}$,
65~times smaller than the most stringent previous limit. 
\end{abstract}
\maketitle

\section{Introduction}

Recent progress in the performance of optical clocks, based on cold
atom ensembles or single ions \citep{Derevianko2011,Poli2013,Ludlow2015,Hong2017},
has become possible due to a strong improvement of the short-term
frequency stability of the ``clock'' lasers that interrogate the
atomic transitions \citep{Kessler2012a,Haefner2015}. Today's clock
lasers are realized by using macroscopic solid-state references, high-finesse
optical cavities. Ultimately, the frequency stability of a clock laser
corresponds to the stability of the cavity's length. The latter is
fundamentally limited by thermal noise, and often also by vibrations
induced by environmental noise. Nevertheless, excellent frequency
instabilities have been reached, with lowest values currently at $4\times10^{-17}$
fractionally \citep{Matei2017}. As of today, only very few approaches
have been identified that have the potential of surpassing the resonator
approach \citep{Meiser2009}. 

One solution is the use of persistent spectral holes in rare-earth-doped
crystals at cryogenic temperatures. Here, a large ensemble of atomic
ions ($\simeq10^{20}$) is embedded in a crystalline host; on the
order of $10^{14}$ ions contribute to a single spectral hole, and
provide a narrow-linewidth frequency reference. This approach,\textcolor{green}{{}
}suggested already more than two decades ago \citep{Macfarlane1987,Sellars1994,Sellin1999,Sellin2001,Pryde2002},
combines advantageous features of atomic and macroscopic solid-state
references.

The first report of frequency-stabilization of a laser to a SH, using
the Pound-Drever-Hall technique, dates back to 1999 \citep{Sellin1999}.
A complication in this approach is that a SH is not a ``static''
reference but is modified dynamically by the interrogating laser field
\citep{Julsgaard2007}, a fact that must be taken into account in
the experimental scheme. More recent work showed that SHs with narrow,
kHz-level linewidths persist for times up to weeks with a high signal
contrast \citep{Thorpe2011}. In order to minimize the modification
of the SHs by the laser radiation to be stabilized, Cook et al. have
developed a technique that uses a pattern of hundreds of SH for frequency
stabilization \citep{Cook2015}. An excellent frequency instability
at the $1\times10^{-16}$ fractional level was thereby achieved for
a 580~nm laser resonant with the ${\rm F}_{0}\rightarrow{\rm D_{0}}$
transition of the Eu\textsuperscript{{\footnotesize{}3+}}:Y\textsubscript{{\footnotesize{}2}}SiO\textsubscript{{\footnotesize{}5}}system
at 580~nm. Recently, in the same system, a heterodyne detection technique
for laser frequency locking to a single spectral hole was demonstrated
\citep{Gobron2017}. Both techniques allow continuous (uninterrupted)
frequency stabilization of a laser.

The long-term frequency stability of a SH on timescales of minutes
and longer and therefore its utility for long-term frequency stabilization
of lasers or for potential studies of fundamental physics depends
also on systematic effects caused by external disturbances. Disturbances
such as variations of temperature and of magnetic field, vibrations,
but also energy exchange on the atomic scale, result in a SH frequency
shift, SH contrast decrease and SH linewidth increase over time. These
issues were studied in detail in the system Eu\textsuperscript{{\footnotesize{}3+}}:Y\textsubscript{{\footnotesize{}2}}SiO\textsubscript{{\footnotesize{}5}}
at 4~K \citep{Thorpe2011,Thorpe2013,Cook2015} and at 3~K \citep{Chen2011}%
, including accurate determinations of the temperature dependence
of the SH frequency. To minimize the influence of this latter effect,
{} \citep{Thorpe2011} implemented a compensation system making use
of the dependence of SH frequency on gas pressure.

In laser frequency stabilization on long time intervals, an extreme
case is the time scale of days and weeks after burning SHs. On this
time scale the drift of the SH frequency and the modification of the
SH lineshape was first investigated in \citep{Chen2011}. A drift
consistent with zero ($<1.5\times10^{-17}/\mathrm{{\rm s}}$ at $1\,\sigma$
level) was observed at 3~K, which was later confirmed in \citep{Leibrandt2013}.In
this paper we extend the study of the long\textendash term properties
of persistent SHs to lower temperature, near 1.2~K. This is one of
the lowest temperatures used so far in SH studies. One particular
aspect of our approach was to probe the spectral holes cautiously.
For example, SHs were left unperturbed spectrally for up to 49 days
until interrogation occurred and then interrogated only once, by a
single absorption scan. 

\newpage
\section{Experiment\label{sec:Experimental-setup}}

\begin{figure}[H]
\centering{}\includegraphics[width=1\columnwidth]{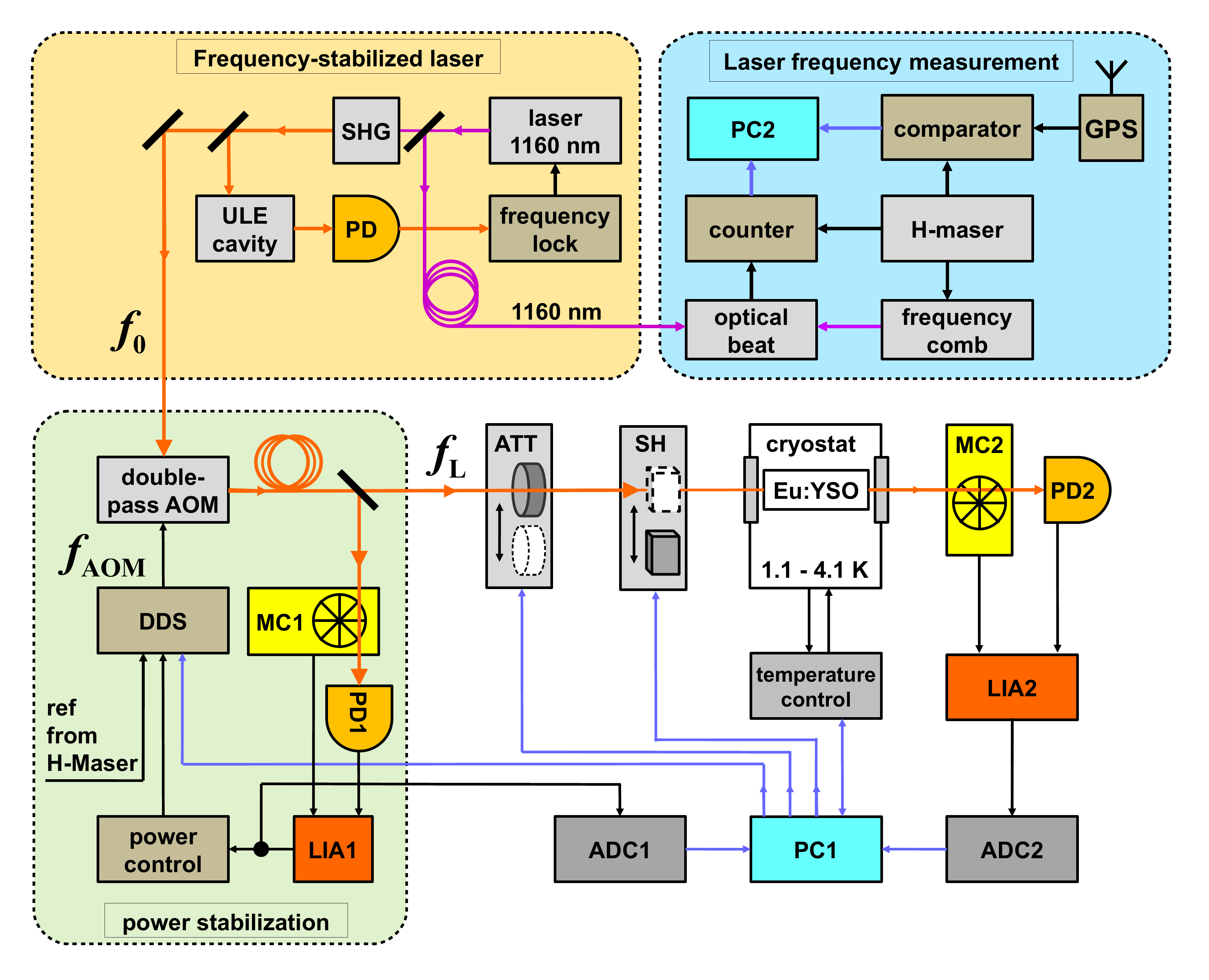}\caption{\label{fig:Setup} Experimental setup. PD - photo detector, SHG -
second harmonic generation, ULE - high-finesse ultra-low thermal expansion
glass optical cavity, AOM - acusto-optic modulator, DDS - direct digital
synthesizer, MC - mechanical chopper, LIA - lock-in amplifier, ATT
- attenuator, SH - shutter, ADC - analog-to-digital converter, PC
- personal computer.}
\end{figure}

Absorption spectroscopy experiments were carried out on the \textsuperscript{7}F\textsubscript{0}$\rightarrow$\textsuperscript{5}D\textsubscript{0}
transition of Eu\textsuperscript{3+} ions of the crystallographic
site~1 in a Y\textsubscript{{\footnotesize{}2}}SiO\textsubscript{{\footnotesize{}5}}
host crystal \citep{Macfarlane1987}, in the temperature range 1.15
- 4.1~K. A schematic of the experimental setup is shown in Fig.~\ref{fig:Setup}.

The Eu\textsuperscript{{\footnotesize{}3+}}:Y\textsubscript{{\footnotesize{}2}}SiO\textsubscript{{\footnotesize{}5}}
crystal with 0.5\% rare-earth ion concentration has dimensions of
$5\times5.5\times10$~mm\textsuperscript{3}, with the polished $5\times5.5$~mm\textsuperscript{2}
facets parallel to the D1 and D2 axes. The crystal was placed inside
a ring-shaped rare-earth permanent magnet with a maximum magnetic
field of approximately 0.8~T. The crystal and magnet were mounted
on a copper plate inside the cryostat. A closed-cycle pulse-tube cooler
cryostat equipped with a Joule-Thomson stage was used for cooling.
This allowed achieving crystal temperatures as low as 1.15~K.

For burning and interrogating SHs we used an external-cavity diode
laser stabilized to a high-finesse ULE optical resonator. A description
of this system, operating at 1156~nm, is given in \citep{Vogt2011a}.
In the present experiment, with the laser operating at 1160~nm, the
linewidth was approximately 50 Hz. The radiation was frequency doubled
to 580~nm, led to the cryostat using a 5~m long polarization-maintaining
optical fiber, and focused into the crystal using a fiber collimator.
The beam diameter in the focus was 300~$\mu$m. The light transmitted
through the crystal was detected with a low-noise silicon photodetector
(PD2). A remotely controlled attenuator, based on a neutral density
filter (ATT) and a mechanical shutter (ST), were used to control the
duration and laser power of SH burning and SH read-out phases.

Before entering the cryostat, the laser light passed a beam splitter
which deflected 30\% of the radiation to a reference Si photodetector
(PD1). Its signal was used for active stabilization of the laser power
entering the crystal. To this end, an analog servo regulated the amplitude
of the RF driver of an acusto-optic modulator (AOM). In addition,
during the spectroscopy experiments, the output of PD1 was used for
normalization of the value of the transmitted laser power, so as to
reduce the influence of the laser power fluctuations on the measurements.
 The laser waves reaching the photodetectors PD1 and PD2 were modulated
with chopper wheels (MC1, MC2) at frequencies of 600~Hz and 650~Hz,
respectively. The photodetector signals were demodulated by respective
lock-in amplifiers (LIA1 and LIA2). A typical lock-in integration
time constant was 0.3~s. The output signals of both lock-in amplifiers
were read out with a 10-bit A/D converter.

To tune the laser frequency, we used the AOM in a double-pass configuration
driven by a computer-controlled DDS. It was referenced to a 10~MHz
reference signal coming from an active hydrogen maser. The complete
experiment including the DDS, the \foreignlanguage{american}{choppers},
lock-in detectors, etc. was computer-controlled with a LabVIEW program.
The absolute frequency $f_{{\rm L}}$ of the wave interrogating the
Eu\textsuperscript{3+}-ions at 580~nm is determined by the stabilized,
but slowly drifting laser frequency at 580~nm, $f_{\mathrm{0}}$,
plus the total frequency shift introduced by the AOM, $f_{{\rm L}}=f_{\mathrm{0}}+2\,f_{{\rm AOM}}$. 

The laser frequency was measured at 1160~nm ($f_{\mathrm{0}}/2$)
relative to the frequency of the active hydrogen maser, using a commercial
erbium-doped fiber laser frequency comb (FC), optically stabilized
to a home-built ULE cavity stabilized 1562~nm laser system by controlling
the FC repetition rate. To this end, a fraction of the $f_{0}/2$
laser radiation was led to the comb laboratory via a 150~m long,
unstabilized fiber. The heterodyne beat signal of the laser with a
comb mode was measured with a dead-time-free frequency counter. The
laser frequency $f_{\mathrm{0}}/2$ was then evaluated in a conventional
way from the beat frequency, the repetition rate, and the carrier
envelope offset frequency. All respective counters were referenced
to the maser. In addition, the maser's long-term frequency drift was
monitored by a GPS receiver. 

\newpage
\section{Spectral holes}

A basic investigation consisted in the determination of conditions
under which the burning produced SHs of appreciable strength but still
having narrow linewidth. To this end, we measured the dependence of
the SH linewidth and SH contrast on the burn phase duration. We define
the SH contrast as $C_{{\rm h}}=(S_{{\rm PD2}}(\Delta=0)-S_{{\rm PD2}}(|\Delta|\gg{\rm FWHM}))/S_{{\rm PD2,max}}$,
where $S_{{\rm PD2}}(\Delta)$ is the laser power measured by the
transmission detector PD2 for a given interrogation laser detuning
from SH line center, $\Delta$. $S_{{\rm PD2,max}}$ is the power
detected on-resonance for a ``deeply'' burnt SH, the crystal then
being essentially transparent.

\begin{figure}[H]
\begin{centering}
\includegraphics[width=0.8\columnwidth]{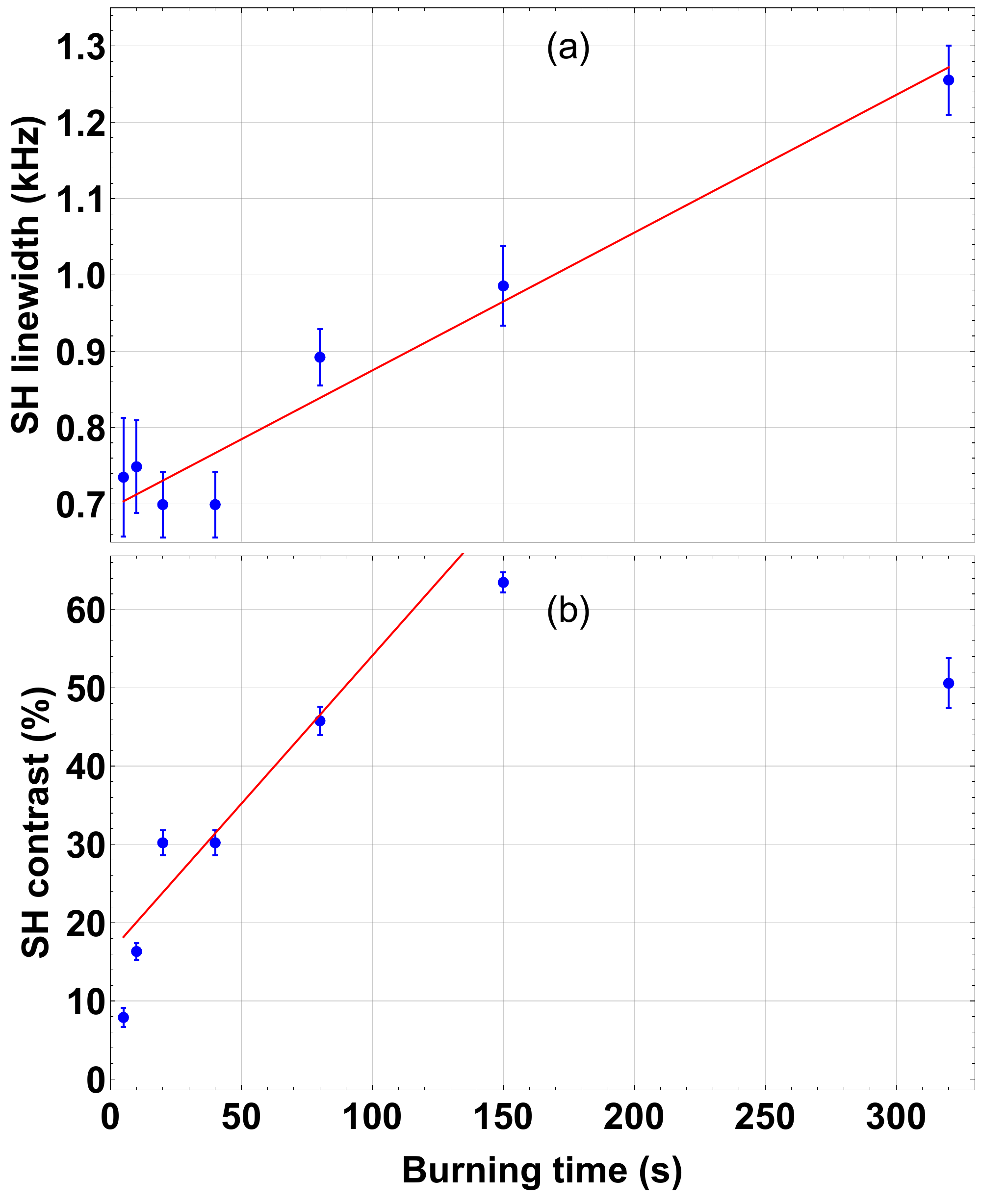}
\par\end{centering}
\caption{\label{fig:FWHM_burntime}a) FWHM linewidths of spectral holes as
a function of the burning time. Burn power is 1.7~nW, beam diameter
is 300~$\mu$m. b) Contrast $C_{\mathrm{h}}$ of the SHs, for different
burn times. }
\end{figure}

Different sets of 5~SHs with a frequency spacing of 20~kHz were
burned with a laser power of 1.7~nW for durations from 10~s to
320~s. Spectroscopy of the holes was carried out by tuning the frequency
$f_{{\rm L}}$ in steps of 100~Hz within a $\pm10\,{\rm kHz}$ frequency
range around the SH center frequencies with a dwell time of 0.7~s.
For these measurements the laser power was reduced to 0.86~nW. The
transmission data taken during such scans were fitted with Lorentzians
and the FWHM (full width at half maximum) linewidth values were determined.
The mean FWHM values of 5-hole sets are presented in Fig.~\ref{fig:FWHM_burntime}a.
The SH linewidth dependence on burn duration was found to be approximately
linear, with a slope of $s=0.10\,\text{kHz/min}$. For small burn
time up to 150~s dwell time, the SH contrast (Fig.~\ref{fig:FWHM_burntime}b)
follows a linear dependence with a slope of 25\%/min. At 320~s burn
time we observed a strong saturation effect. 

\begin{figure}[H]
\begin{centering}
\includegraphics[width=0.8\columnwidth]{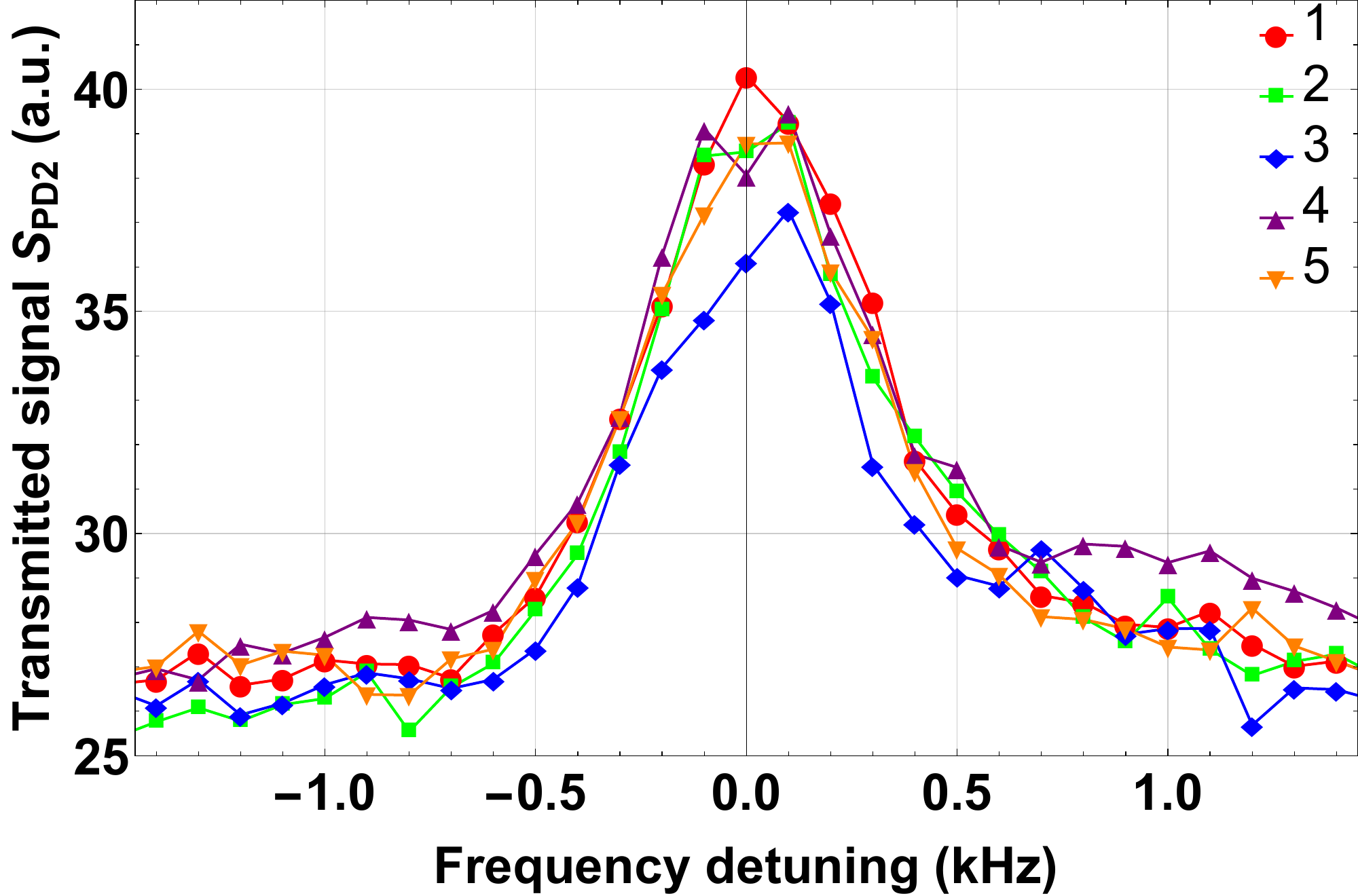}
\par\end{centering}
\caption{\label{fig:TheTypical-absorption-minimal} Transmission signals of
5 different SHs exhibiting the smallest observed linewidths. The frequency
scales for each hole were shifted so as to overlap the peaks. The
measurement conditions are described in the text. }
\end{figure}

We investigated the operational parameters for obtaining minimum
SH linewidths. For example, we produced SHs with the same burn power
level of 1.7~nW and a short burn time of 7~s. The parameters for
the subsequent interrogation scan were optimized to obtain a high
signal-to-noise ratio. The frequency of the interrogation wave was
stepped in 100~Hz increments every 0.5~s.  The SH spectra obtained
are shown in Fig.~\ref{fig:TheTypical-absorption-minimal}. The mean
FWHM linewidth is $0.61(7)$~kHz. This should be compared to the
minimum possible value, the homogeneous linewidth of the SHs, $\Gamma_{{\rm hom,min}}=122\,{\rm Hz}$,
derived from photon echo decay measurements \citep{Yano1991,Equall1994}.
We note that even for our comparatively low linewidth value, the signal
contrast is still appreciable. E.g. for the spectral hole~1 in Fig.~\ref{fig:TheTypical-absorption-minimal},
it was 15\%. Further reduction of the burning laser power and of the
interrogation time did not allow obtaining narrower linewidths. 

\newpage
\section{Temperature-induced frequency shifts and thermal hysteresis\label{sec:Temperature-induced-frequency}}

The temperature-induced shift of the SH center frequency was measured
previously by Könz et al. in the range 4 to 320~K for sites 1 and
2, and a discussion of the underlying physics was given \citep{Koenz2003}.
A precision measurement was performed by Chen et al. \citep{Chen2011}
in the range 3 to 4~K for site~1, and by Thorpe et al. for sites
1 and 2 in the ranges 2.2 - 8.5~K \citep{Thorpe2011} and 2.5 - 5.5~K
\citep{Thorpe2013}. In the present experiment we measured the site~1
shift in the range of 1.15~K to 4.1~K. The determination of the
shift at the lower end of this range is challenging, due to its drop-off
according to an expected $T^{4}$- dependence.The frequency shift
upon temperature cycling was studied by \citep{Chen2011}. No effect
was observed at the few kHz level. Here we set a more precise upper
limit.

\subsection{Procedures\label{subsec:Procedures}}

After starting the laser frequency ($f_{\mathrm{0}}/2$) measurement,
66~SHs were burnt at 1.15~K, spaced by $\delta f_{{\rm L}}=100\,$kHz.
The burning time was set to 2 s, the laser power to 26~nW. The holeburning
procedure had a total duration of about 8~minutes. During the burn
phase the laser frequency exhibited a non-linear drift. Its time dependence
was therefore fitted by a non-linear function. By this procedure we
assigned a center frequency $f_{\mathrm{SH,j}}^{\mathrm{burn}}=f_{\mathrm{SH}}(t_{\mathrm{j}}^{\mathrm{burn}})$
to every SH $j$ created at a particular time $t_{\mathrm{j}}^{\mathrm{burn}}$.
The typical uncertainty was $\sigma_{\mathrm{f_{L},burn}}=60$~Hz.

Over the course of 1~hour, the temperature was increased to 4.1~K
in steps of several 0.1~K and afterwards decreased again. After each
step, at constant temperature, three~so far not interrogated SHs
were scanned and the mean of their center frequencies was determined.
For this SH spectroscopy, the laser power was reduced by a factor
of 10 and the AOM frequency was changed in steps of 100 Hz with a
dwell time of 0.7 s. The recorded data were fitted assuming a Lorentzian
line shape, determining $f_{{\rm fit}}$ and introducing an error
$\sigma_{{\rm fit}}$ of about 10~Hz. 

The frequency of the laser during the spectroscopy was evaluated at
the time instant when the SH center was reached, using a nonlinear
fit as above. This resulted in the SH frequency $f_{\mathrm{SH,j}}^{\mathrm{scan}}$.
Its uncertainty arises from $\sigma_{\mathrm{fit}}$ and from the
error of the FC measurement $\sigma_{\mathrm{f_{L},scan}}$. Finally,
the frequency shift of a SH $j$ is $\Delta f_{\mathrm{SH,j}}=f_{\mathrm{SH,j}}^{\mathrm{scan}}-f_{\mathrm{SH,j}}^{\mathrm{burn}}.$
We assigned to it the uncertainty $\sigma_{{\rm shift}}=\sqrt{\sigma_{\mathrm{f_{L},burn}}^{2}+\sigma_{\mathrm{f_{L},scan}}^{2}+\sigma_{\mathrm{fit}}^{2}}$.
A typical value was 90~Hz.

\begin{figure}[H]
\begin{centering}
\includegraphics[width=1\columnwidth]{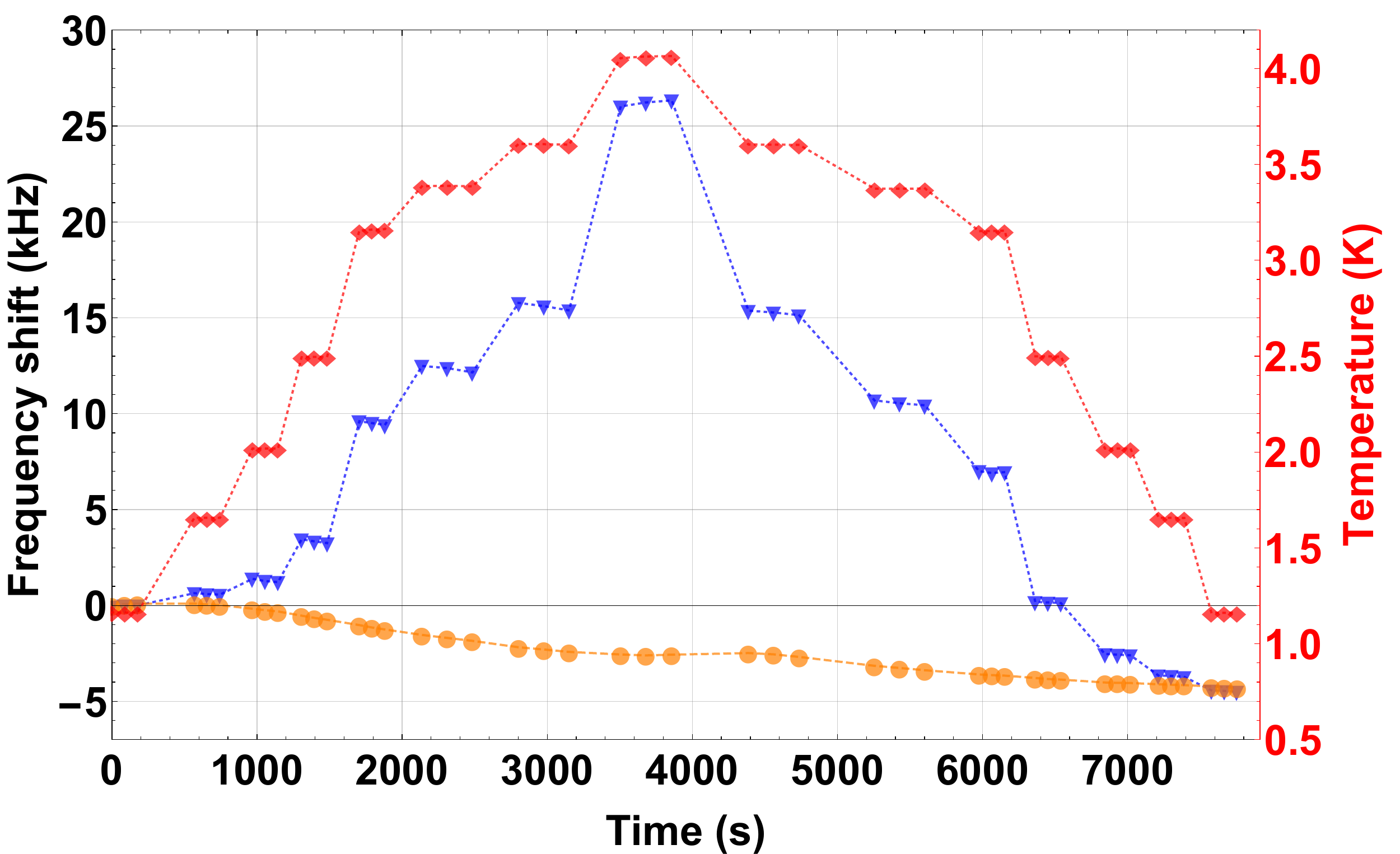}
\par\end{centering}
\caption{\label{fig:Hysterese}Variation of the SH center frequencies as found
from the AOM frequency shift $2\,\Delta f_{\mathrm{{\rm AOM,}j}}(t)$,
(blue down-pointing triangles) as the crystal temperature (red diamonds)
is increased from 1.15~K to 4~K and cooled back again. The independently
occurring drift of the frequency-doubled laser frequency, $\bigtriangleup f_{\mathrm{0}}(t)$,
is shown as orange circles. Each blue point corresponds to one scan
over one SH.}
\end{figure}

An overview of the recorded (raw) data is presented in Fig.~\ref{fig:Hysterese}.
The red line shows the temperature measured at the crystal. The orange
line shows explicitly the drift of the frequency-doubled laser frequency
$\Delta f_{\mathrm{0}}(t)$%
. The blue line shows $2\,\Delta f_{\mathrm{AOM,j}}(t)$ which is
the difference of the frequency offsets produced by the AOM at the
observation time $t$ and at the burning time.

\subsection{Thermal frequency shift}

\begin{figure}[H]
\begin{centering}
\includegraphics[width=0.8\columnwidth]{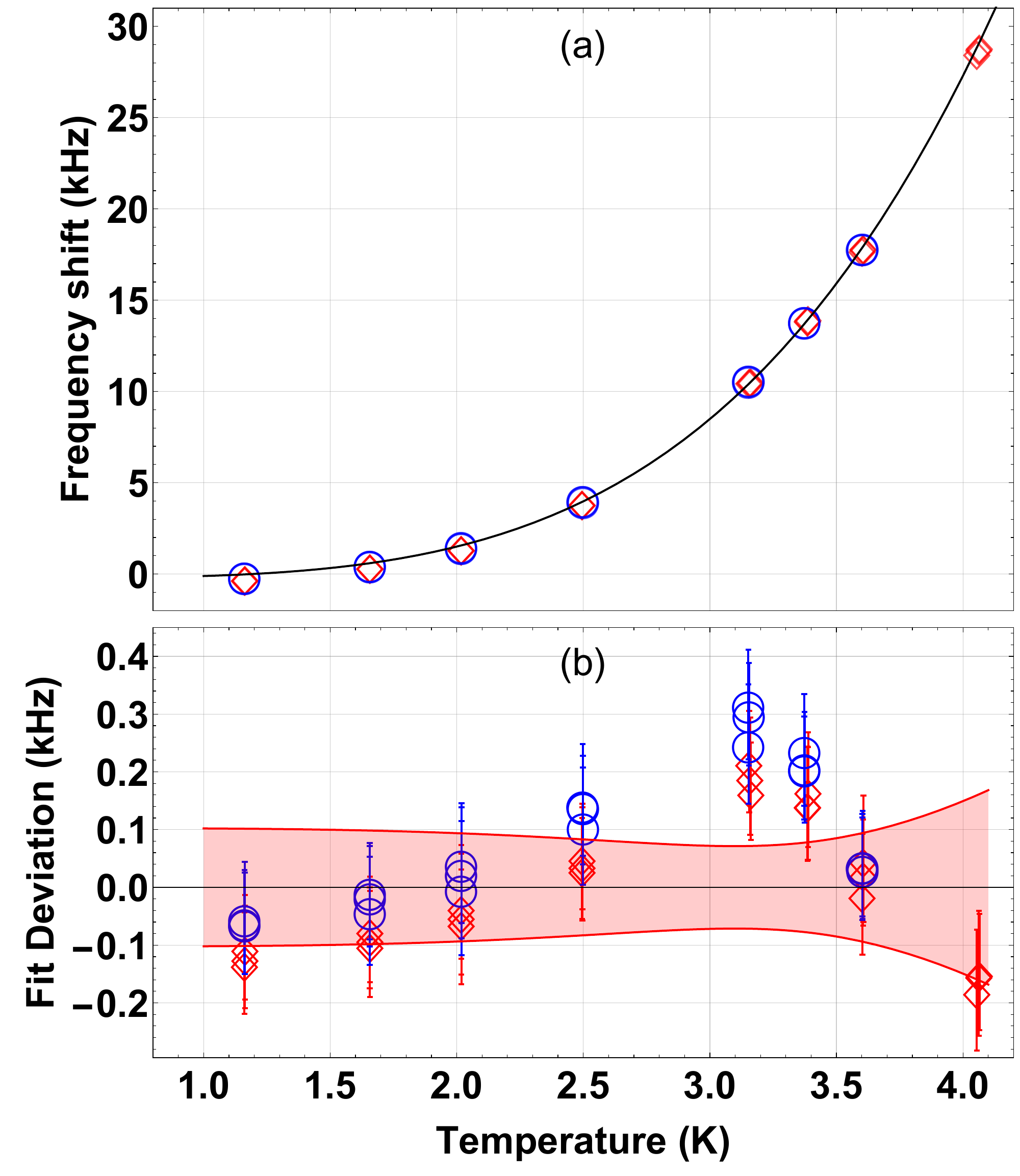}
\par\end{centering}
\caption{Temperature dependence of the SH center frequencies for rising (red
diamonds) and falling (blue circles) temperature. The black line in
the upper graph is a fit to the model function $A\,T\text{\textsuperscript{4}}+B$.
The lower graph b) shows the deviations from the fit and the $\pm1\sigma$
fit uncertainty interval. }
\label{figTdep}
\end{figure}

The frequency shift data is sheown as a function of temperatumeasuredre
in Fig. \ref{figTdep}~(a).A theoretical model of the SH shift predicts
a $T^{4}$ dependence, caused by a two-phonon Raman process affecting
the impurity ions \citep{Macfarlane1987,McCumber1963,Koenz2003}.
Our data was therefore fitted with the function $A\,T^{4}+B$. This
resulted in an accurate description of the data, and small residuals,
see panels (a) and (b) in Fig.~\ref{figTdep}, with $A=108(0.3)\,{\rm Hz/K^{4}}$
and $B=-216(37)\,{\rm Hz}$. Note that because of the small hysteresis
(see section \ref{subsec:Thermal-hysteresis}) we take into account
the data for both rising and falling temperatures. In the work of
\citep{Koenz2003} a value of $A'=166(10)$~Hz/${\rm K}^{4}$ has
been found for the temperature induced shift at site 1. However, only
values above 70~K were considered in the determination of the coefficient.

Our coefficient $A$ also differs from the value $76(15)\,{\rm Hz/K^{4}}$
(for site~1) measured previously by Thorpe \emph{et. al} \citep{Thorpe2013}.
As shown in that work the application of external pressure on the
crystal through helium gas induces a linear frequency shift countering
the effect of the temperature induced frequency shift. This could
explain the lower coefficient.

\begin{figure}[H]
\centering{}\includegraphics[width=0.8\columnwidth]{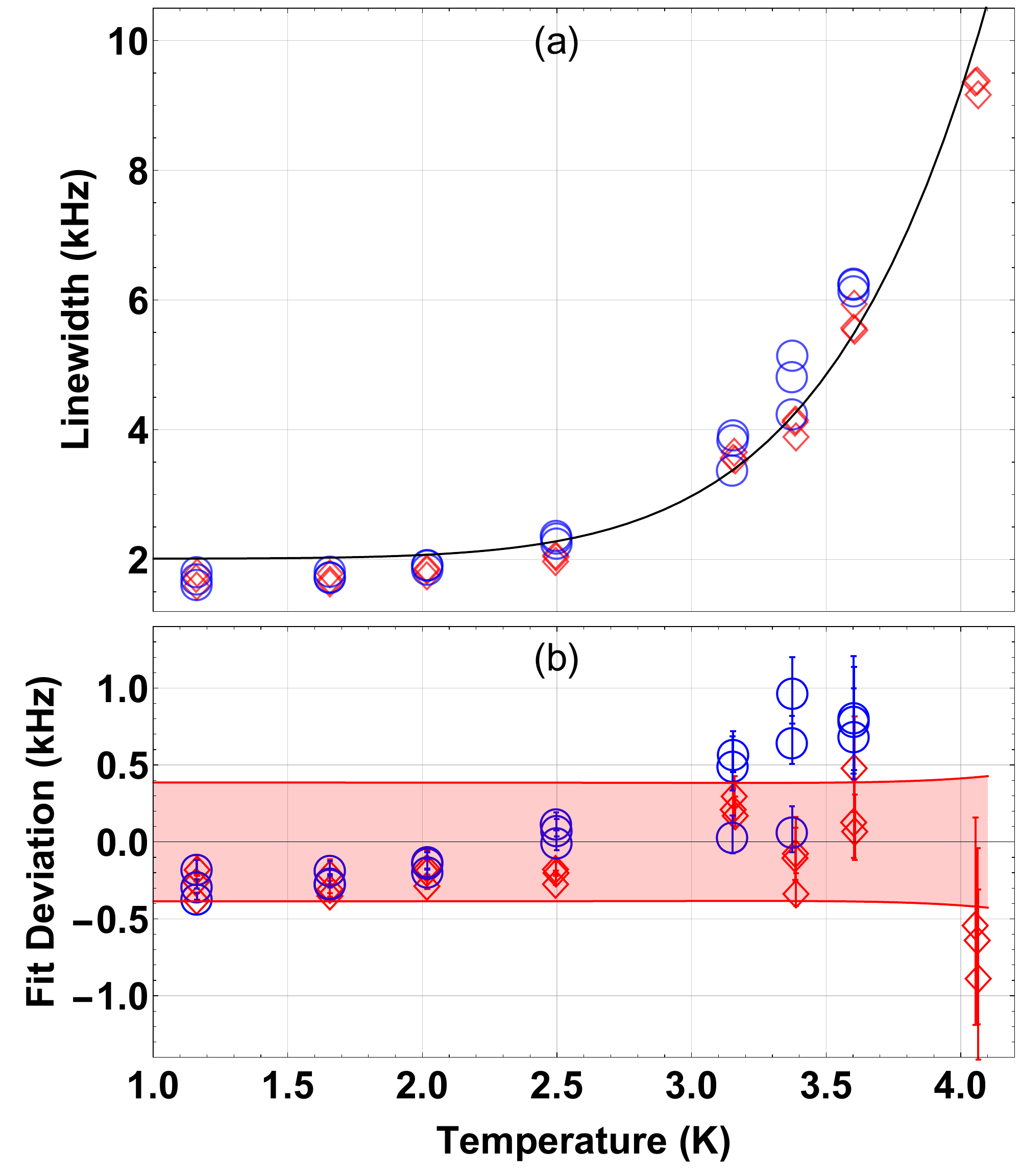}\caption{Temperature dependence of the SH linewidth for rising (red diamonds)
and falling (blue circles) temperature. The black line in the upper
graph (a) is a fit to the model function $C\,T^{7}+D$. The lower
graph (b) shows the deviations from the fit and the $\pm1\sigma$
fit uncertainty interval.\label{fig:Temperature-dependence-of-linewidth}}
\end{figure}

Figure \ref{fig:Temperature-dependence-of-linewidth} shows the temperature
dependence of the SH linewidth. The theoretical model \citep{McCumber1963,Koenz2003}
predicts a dependence of the type $C\,T^{7}+D$. A fit of this function
to our data yields $C=0.44(2)\,{\rm Hz}/{\rm K}^{7}$.
 The value differs significantly from the value found by \citep{Koenz2003},
$7.2\times10^{-4}\,{\rm Hz}/{\rm K}^{7}$. However, this value was
obtained from photon-echo-decay measurements. This method differs
from our method: the photon-echoes were measured on a timescale of
microseconds whereas our measurements where on the timescales of minutes
(the typical duration of a frequency scan over a SH linewidth).

\subsection{Thermal hysteresis\label{subsec:Thermal-hysteresis}}

In Fig.~\ref{fig:Hysterese} the AOM-induced frequency shift (blue
line) and the independently measured laser frequency drift (orange
line) rejoin at the end of the temperature cycle. Correspondingly,
the red and blue data point at $T=1.15$~K in Fig.~\ref{fig:Hysterese}~(b)
nearly coincide. This shows that the thermal hysteresis of the SH
frequencies is very small. More precisely, the residual shift upon
returning to the initial temperature was $\delta f_{\mathrm{hyst}}=$
59~Hz which is within the measurement error $\sigma_{\mathrm{shift}}$.
Relating the sum of $\delta f_{\mathrm{hyst}}$ and $\sigma_{\mathrm{shift}}$
to the maximum frequency shift of 27~kHz at $T=4$~K allows us to
state an upper limit of $6\times10^{-3}$ for the fractional hysteresis.

\newpage
\section{Long-term frequency drift measurement\label{sec:Long-term-frequency-drift}}

\subsection{Properties of long-lived spectral holes}

In order to probe in a careful way the intrinsic long-term stability
of the frequency of persistent SHs, we modified the experimental protocol
compared to our previous work \citep{Chen2011} in two significant
aspects: (1) operation at significantly lower temperature, 1.15~K,
and (2) not performing multiple read-outs of the same SH. Our procedure
consisted in initially burning a sufficiently large ``reservoir''
of 200~SHs with a frequency spacing $\delta f_{{\rm L}}=200\,$kHz.
The burning time was set to 2~s with a relatively high laser power
of 86~nW. The hole-burning procedure had a total duration of about
15 minutes. For the SH spectroscopy, the laser power was reduced by
a factor of 10 and the AOM frequency was changed in steps of 100~Hz
with a dwell time of 0.7~s per frequency value. 

\begin{figure}[H]
\begin{centering}
\includegraphics[width=0.72\columnwidth]{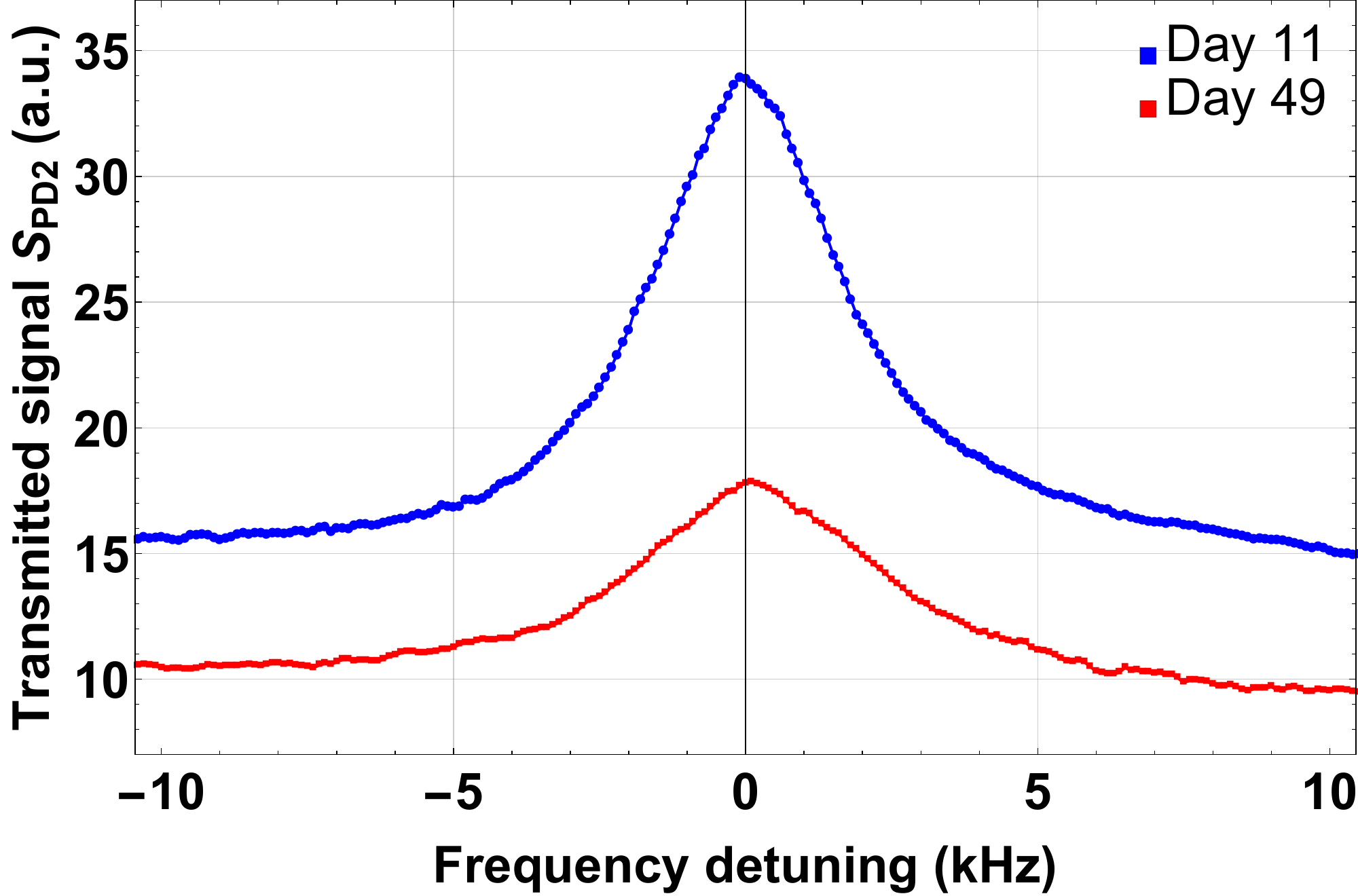} 
\par\end{centering}
\caption{\label{fig:-Typical-spectral} Typical spectral holes observed during
the long-term drift measurement. The upper blue points show a scan
of a spectral hole 11~days after burning. The lower red points show
another SH scanned 49~days after burning. In order to illustrate
the typical deformation of the SHs, the respective frequency scales
were shifted so as to center the peaks.}
\end{figure}

A typical line scan obtained under such conditions is shown in Fig.~\ref{fig:-Typical-spectral}.
During the scan the laser frequency was measured by the FC. The determination
of the SH frequency shift $\Delta f_{\mathrm{SH,j}}$ from $f_{\mathrm{SH,j}}^{\mathrm{scan}}$,
$f_{\mathrm{SH,j}}^{\mathrm{burn}}$ and the measurement errors $\sigma_{\mathrm{f_{L},burn}}$,$\sigma_{\mathrm{f_{L},scan}}$
and $\sigma_{\mathrm{fit}}$ were obtained in the same fashion as
described in sec.~\ref{sec:Temperature-induced-frequency}.

\begin{figure}[H]
\centering{}\includegraphics[width=0.8\columnwidth]{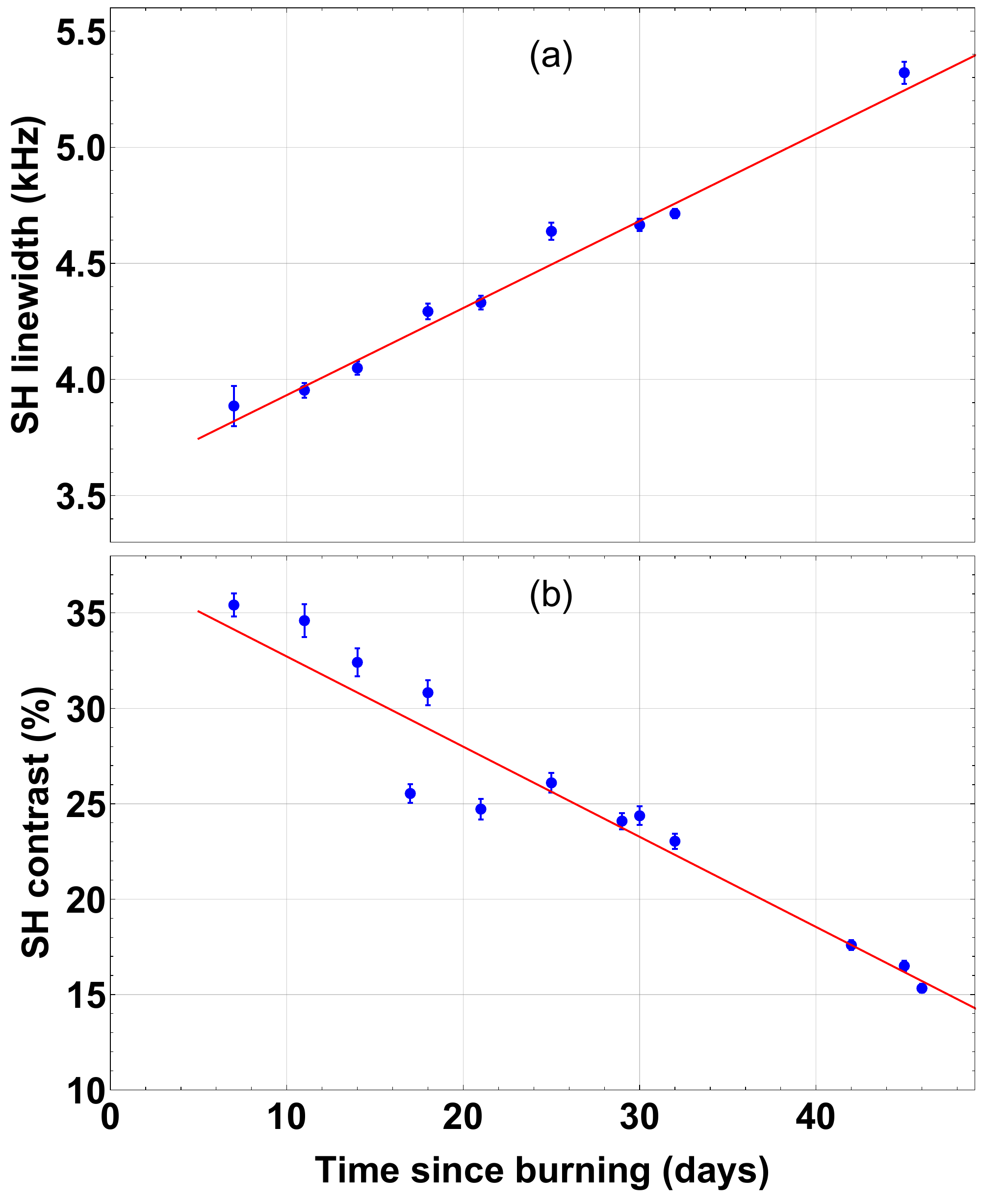}\caption{(a) Long-term variation of the SH linewidth (FWHM) and (b) of the
SH contrast $C_{{\rm h}}$. Each SH was scanned only once.}
\label{figSHDevelopment}
\end{figure}

Several times per week, over a period of 49~days, a subset of new
(previously not interrogated) SHs was scanned. At the same time, the
laser frequency was measured with the FC. \ref{fig:Determinaton of mean SH frequency (green)}
shows an example of the result obtained on a particular day. The long-term
variation of the  SHs linewidths and contrasts are presented in Fig.~\ref{figSHDevelopment}.
After 49~days we observed an increase of the linewidth by about 25\%.
The $1/e$ - SH lifetime, extrapolated from the contrast decrease,
is 50~days. This value is consistent with \citep{Koenz2003} who
measured a value $\simeq23$~days at 2~K for site~1.

\subsection{Long-term frequency drift}

\begin{figure}[H]
\centering{}\includegraphics[width=0.8\columnwidth]{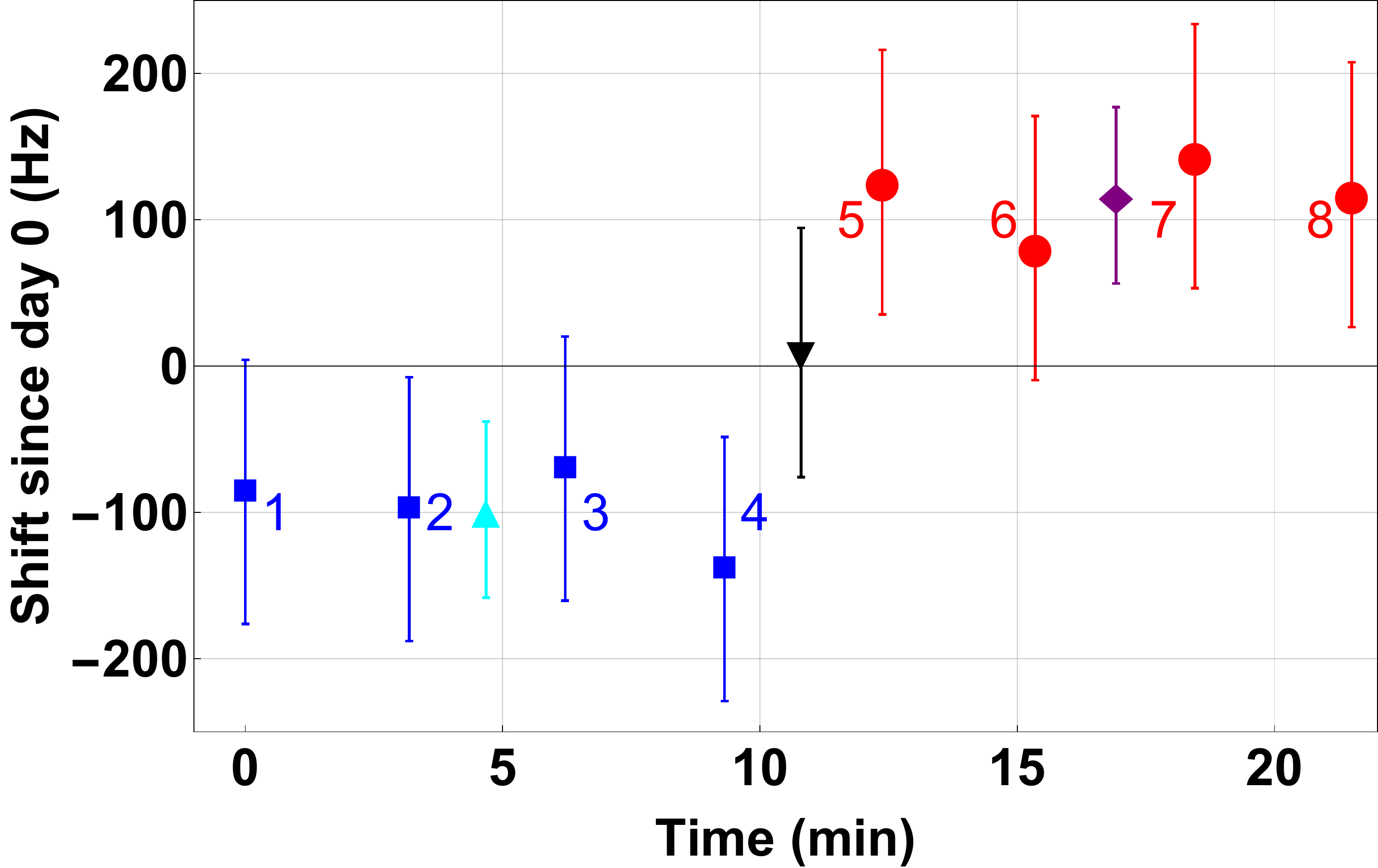}\caption{\label{fig:Determinaton of mean SH frequency (green)} Determination
of the mean SH frequency $f_{\mathrm{shift}}(t)$ (black down-pointing
triangle) on day~18 of the long-term measurement. Four SHs ($j=$1
to 4, blue squares) were scanned once with decreasing frequency and
another set ($j=$5 to 8, red circles) was scanned once with increasing
AOM frequency. The mean value of the ``decreasing'' scans $\langle\Delta f_{\mathrm{SH,j}}\rangle_{\mathrm{dec}}$
is shown as cyan up-pointing triangle, of the ``increasing'' scans
$\langle\Delta f_{\mathrm{SH,j}}\rangle_{\mathrm{inc}}$ as purple
diamond. }
\end{figure}

A systematic frequency shift occurs when determining the SH center
frequency: the latter depends on the sign of the frequency introduced
by the AOM during the line scan. We measured this effect precisely
for the above mentioned settings by burning 20~SHs within 2~minutes
and scanning them 2 to 18~minutes later. One half of the SHs was
scanned by increasing $f_{{\rm AOM}}$; the other half was scanned
in reverse. We measured a difference of 298(47)~Hz between the average
center frequencies $\langle\Delta f_{\mathrm{SH,j}}\rangle$ of the
two subsets. Therefore, in our long-term measurements, we determined
the frequency shifts as follows. At a particular (nominal) time $t$,
we measured, within approximately half an hour, the mean SH center
frequency $\langle\Delta f_{\mathrm{SH,j}}\rangle_{\mathrm{inc}}$
of a set of at least 4 SHs scanned with rising AOM frequency and the
mean frequency $\langle\Delta f_{\mathrm{SH,j}}\rangle_{\mathrm{inc}}$
of another set of at least 4 SHs with falling frequency. The average
frequency shift $f_{\mathrm{shift}}(t)$ of the SHs at time $t$ is
the mean of $\langle\Delta f_{\mathrm{SH,j}}\rangle_{\mathrm{dec}}$
and $\langle\Delta f_{\mathrm{SH,j}}\rangle_{\mathrm{inc}}$. An example
of such a determination is shown in Fig.~\ref{fig:Determinaton of mean SH frequency (green)}. 

\begin{figure}[H]
\centering{}\includegraphics[width=0.8\columnwidth]{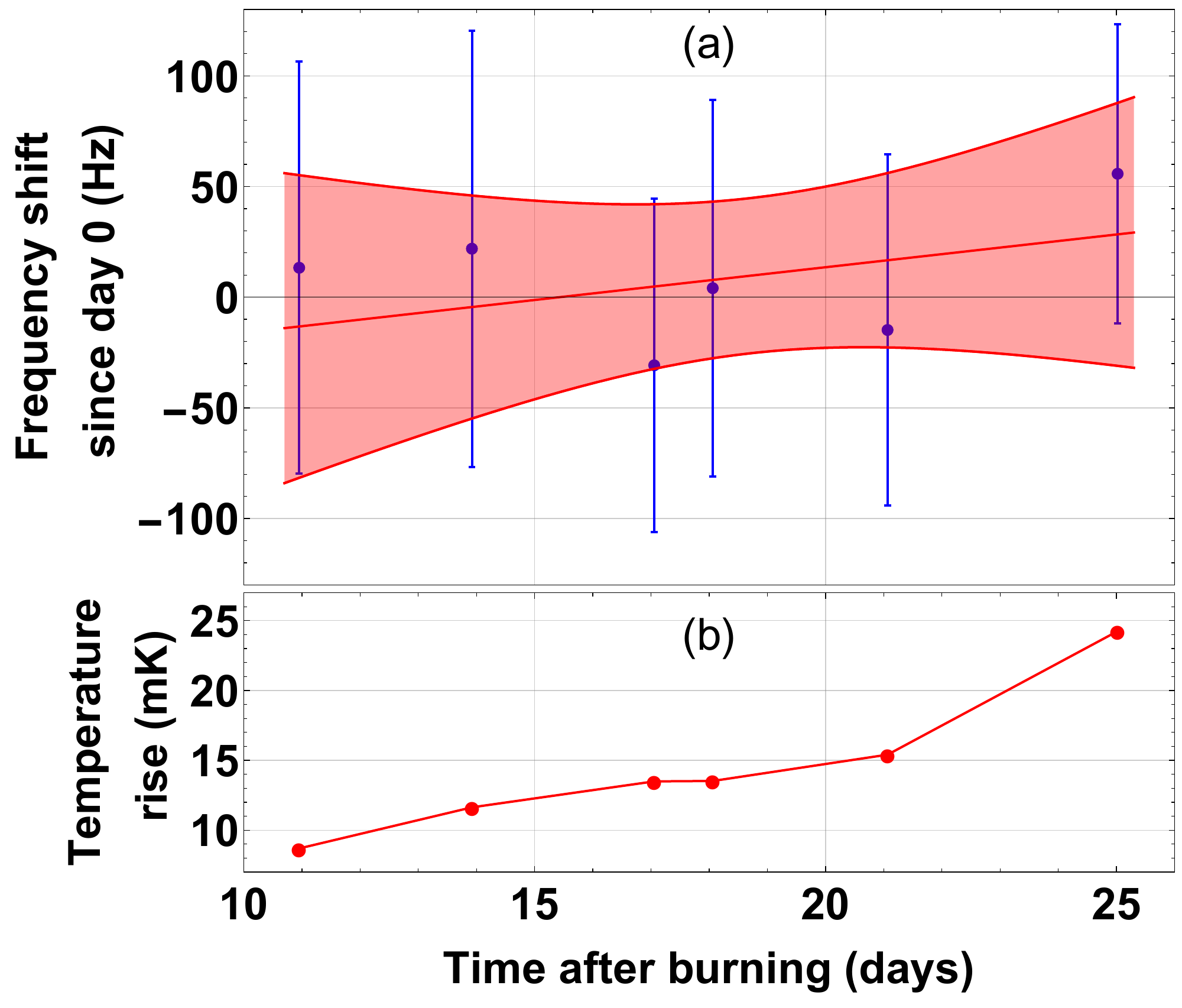}\caption{\label{fig:Long-term drift}(a) The change in SH frequency, $f_{\mathrm{shift}}(t)$
measured over a period of 14~days. All SHs were burned at time $t=0$.
Red line: linear fit with slope 3(7)~Hz/day, or $1.3(10)\times10^{-19}\,{\rm s}^{-1}$
fractionally. The shaded region shows the $\pm1\sigma$ uncertainty
of the fit. (b) Temperature measured at the location of the Eu\protect\textsuperscript{3+}:Y\protect\textsubscript{2}SiO\protect\textsubscript{5}
crystal. A calibrated Cernox temperature sensor having a specified
inaccuracy of 5~mK was used.}
\end{figure}
The complete measurement lasting two weeks is shown in Fig.~\ref{fig:Long-term drift}.
A linear fit of the data shows a fractional SH frequency drift of
$1.3(10)\times10^{-19}/{\rm s}$, where the uncertainty is statistical.
The drift rate is consistent with zero.

The statistical uncertainty originates from the three contributions
already discussed in Sec.~\ref{subsec:Procedures}, which resulted
in $\sigma_{{\rm shift}}\simeq90$~Hz for an individual measurement
$\Delta f_{\mathrm{SH,j}}$. 

A systematic effect is a frequency shift due to long-term temperature
drift. The temperature increased monotonically by 30~mK during the
measurement period, see Fig.~\ref{fig:Long-term drift}~(b). Taking
into account the measured temperature sensitivity (see Sec.~\ref{sec:Temperature-induced-frequency})
the resulting total SH frequency shift is 10~Hz, which is not negligible.
Therefore, the data shown in Fig.~\ref{fig:Long-term drift}~(a)
was corrected for the respective calculated thermal shift.

A further possible systematic error is a drift of our frequency reference,
the hydrogen maser. A comparison of its frequency with the 1~PPS
signal received from GPS showed that the maser frequency drift during
the measurement period was on the order of $1\times10^{-20}/{\rm s}$,
and can thus be neglected

\section{Summary and Conclusion }

We measured with high frequency resolution and high accuracy the linewidth,
the long-term frequency drift and the temperature-induced frequency
shift and line broadening of persistent spectral holes in Eu\textsuperscript{{\footnotesize{}3+}}:Y\textsubscript{{\footnotesize{}2}}SiO\textsubscript{{\footnotesize{}5}}
at a temperature significantly lower than previously, 1.15~K. Our
measurements demonstrated a significant increase of the spectral hole's
lifetime in comparison to results at 3 - 4~K \citep{Chen2011,Cook2015},
confirming the estimations \citep{Koenz2003}. We determined the properties
of spectral holes as long as 49 days after burning and found that
even at that ``age'', the holes still exhibited good signal-to-noise
ratio and a reasonably small linewidth (5.4~kHz), if previously left
undisturbed. No long-term drift of the spectral holes center frequencies
over 14~days could be observed, with an $1\sigma$ upper limit of
$2.3\times10^{-19}/{\rm s}$. This is 65~times smaller than the upper
limit measured in a previous experiment at 3~K, $1.5\times10^{-17}{\rm /s}$
\citep{Chen2011}. For comparison, cryogenic silicon cavities with
drifts as small as $5\times10^{-19}/{\rm s}$ to $1.4\times10^{-20}/{\rm s}$
have been reported \citep{Hagemann2014,Wiens2016}. We expect that
it is feasible to reduce the uncertainty of the spectral hole drift
by increasing the observation time interval, which in our case was
limited because of cryostat performance. A measurement of the temperature-induced
hole line shift over the range 1.2~K - 4.1~K accurately confirmed
the predicted variation with the fourth power of the temperature.
No hysteresis in the hole frequency was found within the measurement
error after heating the crystal from 1.15~K to 4.1~K and cooling
back to 1.15~K. 

The long-term properties of spectral holes in this particular system
do appear to make it suitable as a long-term-stable frequency reference.
In particular, it is favorable that at 1.15~K the temperature sensitivity
is only $\simeq1\times10^{-12}/{\rm K}$ fractionally. This is comparable
to the lowest values achieved with cryogenic silicon cavities \citep{Wiens2016}.
If required for achieving the most demanding performance, an active
temperature stabilization of a YSO crystal to the $\mu$K level could
be implemented \citep{Chen2011} and this could reduce the effects
of temperature instability to the $10^{-18}$-level. 

The SH linewidth observed in this work is also comparable to that
of silicon cavities \citep{Wiens2016}. The signal-to-noise ratio
of the spectral hole signals is, of course, significantly lower than
for a cavity. It could be increased e.g. by using longer crystals,
interrogating several crystals in parallel, and increasing the europium
concentration. 

\textbf{Acknowledgments}

The authors thank D.~Iwaschko for his technical assistance. R.O.
acknowledges a fellowship from the Prof.-W.-Behmenburg-Schenkung.
This work was performed in the framework of project Schi~431/15-1
of the Deutsche Forschungsgemeinschaft.

\bibliographystyle{apsrev}
\bibliography{Manuscript_v10.bbl}  

\end{document}